\documentclass[apl,twocolumn,showpacs,groupedaddress,preprintnumbers]{revtex4}

\usepackage{graphicx}
\usepackage{bm}
\usepackage[mathlines]{lineno}
\usepackage{ulem}
\usepackage{color}
\usepackage{comment}

\begin{document}

\title[Photon Energy Dependence of Kerr Rotation in Chalcogenide Superlattices]{Photon Energy Dependence of Kerr Rotation in Chalcogenide Superlattices}

\author{Takara Suzuki$^1$, Richarj Mondal$^1$, Yuta Saito$^2$, Paul Fons$^2$, Alexander V. Kolobov$^{2,3}$, Junji Tominaga$^2$, Hidemi Shigekawa$^1$, and Muneaki Hase$^1$}
\affiliation{$^{1}$Department of Applied Physics, Faculty of Pure and Applied Sciences, University of Tsukuba, 1-1-1 Tennodai, Tsukuba 305-8573, Japan}
\affiliation{$^{2}$Nanoelectronics Research Institute, National Institute of Advanced Industrial Science and Technology, Tsukuba Central 5, 1-1-1 Higashi, Tsukuba 305-8565, Japan}
\affiliation{$^{3}$Department of Physical Electronics, Faculty of Physics, Herzen State Pedagogical University, St. Petersburg, 191186, Russia}
\email{mhase@bk.tsukuba.ac.jp}
\vspace{10pt}

\begin{abstract}
We report on pump-probe based helicity dependent time-resolved Kerr measurements of chalcogenide superlattices, consisting of alternately stacked GeTe and Sb$_{2}$Te$_{3}$ layers under infrared excitation. The Kerr rotation signal consists of the specular Inverse Faraday effect (SIFE) and the specular optical Kerr effect (SOKE), both of which are found to monotonically increase with decreasing photon energy over a sub-eV energy range. Although the dependence of the SIFE can be attributed to a response function of direct third-order nonlinear susceptibility, the magnitude of the SOKE reflects cascading second-order nonlinear susceptibility resulting from  electronic transitions from bulk valence band to interface-originating Dirac states of the superlattice. 
\end{abstract}

\maketitle

\section{Introduction}

Structures fabricated from the alternate stacking of GeTe and Sb$_{2}$Te$_{3}$ layers, referred to as interfacial phase change memory (iPCM) or chalcogenide superlattices (CSL), are known to form either 
Weyl or Dirac semimetal systems, which have closely related topological insulator (TI) properties, depending on the 
thickness of the constituent layers, and are believed to be excellent candidates for power-saving nonvolatile memory applications \cite{simpson2011,lotnyk2017,kim2017,makino2014}.  {\it Ab initio} band structure simulations of CSL structures show a gapless band intersection, a so-called Dirac cone, is formed for the so-called inverted Petrov phase (Dirac semimetal phase) due to the hybridization of the electronic wave functions across the interfaces between the GeTe and Sb$_{2}$Te$_{3}$ layers \cite{tominaga2014}.
Although recent angle-resolved photoemission spectroscopy (ARPES) study have examined the band structure of the chalcogenide alloy (Ge$_{2}$Sb$_{2}$Te$_{5}$) \cite{Kellner2018}, there has been no direct experimental evidence for a Dirac-like electronic band structure in CSLs. To further examine the topological properties of CSLs and to further spintronics applications, experimental investigation of the band structure of CSLs is required. 

TI-like characteristics can be discerned optically using circularly polarized light due to the fact that the angular momentum of circularly polarized photons can induce spin-selective excitations to/from topological surface states, which are a key signature of topological character \cite{mclver2012,hsieh2011,boschinii2015,Takeno2018}. 
Under excitation by circularly polarized light, some solid materials show helicity dependent optical activity, such as polarization rotations and circular dichroism \cite{qiu2000,tse2010,wilks2004a,wilks2003}. 
In this study, the authors focus on nonlinear optical effects induced by femtosecond laser irradiation, these include the inverse Faraday effect (IFE) \cite{ziel1965,kimel2005} and the optical Kerr effect (OKE) \cite{shen1984}, both of which lead to polarization rotation. 
The former rotation occurs due to magnetization induced by the circularly polarized light irradiation while the latter is a result of birefringence through a third-order nonlinear optical process governed by the nonlinear susceptibility $\chi ^{(3)}$.

In a pump-probe experiment, the polarization rotation of the incident probe beam in a material is sensitive to the helicity of excitation pulse. 
For instance, simple metals \cite{wilks2004}, magnetic materials \cite{kimel2005} or chalcogenide compounds \cite{mondal2018,mondal2018a} have shown specular IFE (SIFE) or specular OKE (SOKE). 
It is noteworthy that, in CSL, the SOKE signal becomes larger than the SIFE even though the SIFE is usually larger in normal insulators such as chalcogenide alloys \cite{mondal2018a}. This is believed to arise from the large $\chi ^{(2)}$ of the topological Dirac state since third-order nonlinear optical processes are constituted from not only direct third order processes but also cascading second-order nonlinear processes as expressed by $\chi ^{(2)} \cdot \chi ^{(2)}$ \cite{mondal2018,mondal2018a}.
In a previous experiment by Mondal {\it et al.}, only near-infrared 830 nm femtosecond laser pulses were used \cite{mondal2018a}. 
To explore the relationship between the SOKE (SIFE) signal and the band structure of CSLs, pump-probe measurements over a wide wavelength (or photon energy) range are required. 
Here, we report on the photon energy dependence of the SIFE and SOKE signals in CSL by observing the polarization rotation of the linearly polarized probe beam. In particular, 1200 $\sim$ 1600 nm IR pulses were used to investigate the TI band structure in detail. We found that both the SIFE and SOKE signals were strongly enhanced when the photon energy approached the energy required for direct excitation from core valence band states to topological Dirac states \cite{tominaga2014,mondal2018b,Krizman2018}.  

\section{Experimental}

\begin{figure}
\centering\includegraphics[width=80mm]{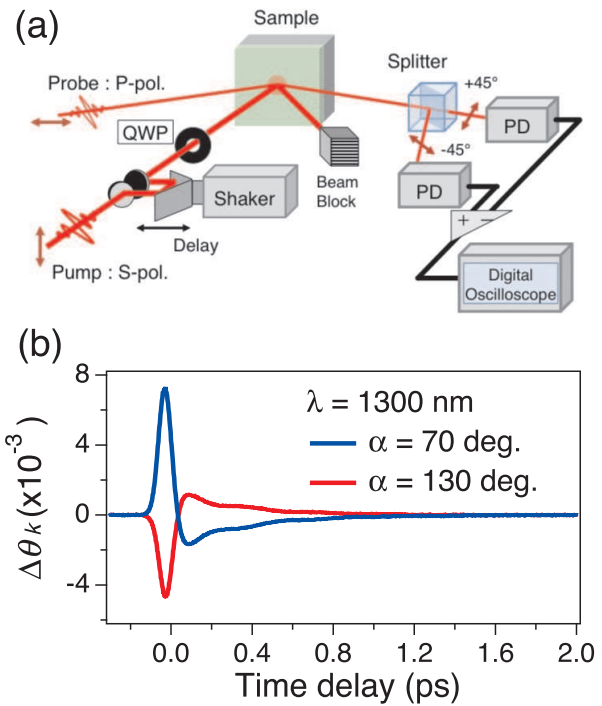}
\caption{(a) Optical setup for transient Kerr rotation measurement is illustrated. (b) The time resolved Kerr rotation data obtained at 1300 nm. Strong peak was observed at zero time delay. Left ($\alpha$ = 70$^{\circ}$) and right-handed elliptically polarized light excitation ($\alpha$ = 130$^{\circ}$) results in different peak sign ($\pm$). 
}
\label{fig:setup}
\end{figure}

Reflection geometry pump-probe measurement were carried out to obtain the transient Kerr rotation of the reflected probe pulse at room temperature \cite{mondal2018}. 
The reflected probe pulse was decomposed into two beams polarized at $+45^{\circ}$ and $-45^{\circ}$, respectively, by a Wollaston prism and detected by balanced Si-PIN photo-detectors as illustrated in Fig. \ref{fig:setup}(a). The intensity difference of the two beams was defined as the Kerr rotation of the probe polarization: $\Delta\theta_{k}$. Femtosecond laser pulses (repetition rate, 80 MHz; pulse width, 20 fs; and average power, 525 mW) were generated by a mode locked Ti$^{3+}$: sapphire laser oscillator. Its output was amplified by a regenerative amplifier system (RegA9040) to produce pulses of 40 fs duration with an average power of 500 mW at a 100 kHz repetition rate. Finally, the amplified pulses were sent to an optical parametric amplifier (OPA9580) to generate IR pulses, whose wavelength was tuned from 1200 to 1600 nm (1.03 -- 0.775 eV). The 600 nm visible photons were also generated by doubling the 1200 nm photon energy using second harmonic generation with a 1 mm-thick $\beta$-barium borate (BBO) crystal. 
The pump and probe wavelengths were identical in the present experiment. The output beam was split into an intense pump and weak probe beams.  
The pump and the probe beam were co-focused onto the sample to a spot size of about 70 $\mu$m with an incident angle of about $15^{\circ}$ and $10^{\circ}$ with respect to the sample normal, respectively. 
A quarter wave plate (QWP) was inserted in the pump optical path to allow the pump helicity to be varied continuously from linear to circular, and data was recorded for every 10-degree rotation of the QWP. Analyzing the optical response as a function of the pump helicity offers insight into the role of topological Dirac state in the optical response. The Kerr rotation was recorded as a function of time delay between the pump and probe over a range of 20 ps by means of a horizontally shaking mirror with a scanning frequency of 10 Hz \cite{Hase2012,Hase2015}. 
The pump power was kept constant at 20 mW over the wavelength range used, which corresponds to a fluence of $\approx$ 16 mJ/cm$^{2}$, a value calculated using a focused beam diameter of $\approx$ 70 $\mu$m (measured using a knife-edge). 
An optical penetration depth of $\approx$ 100 nm at 1550 nm was calculated from the absorption coefficient \cite{lee2005}, suggesting negligibly small inhomogeneous excitation effects along the sample depth. 

The (GeTe)$_{2}$/Sb$_{2}$Te$_{3}$ sample used was fabricated by radio-frequency magnetron sputtering at 230$^{\circ}$C and was deposited onto a Si (100) substrate. For better z-axis orientation 
\cite{saito2015,saito2016}, the Sb$_{2}$Te$_{3}$ layers were grown first to a thickness of 3 nm at room temperature and subsequently GeTe (0.8 nm) and Sb$_{2}$Te$_{3}$ (1.0 nm) layers were  alternately deposited for 20 cycles ($\sim$36 nm) at elevated temperature. To prevent oxidization, a 20-nm-thick ZnS-SiO$_{2}$ layer was deposited on top of the as-grown CSL without breaking the vacuum.

\section{Results and discussion}
\begin{figure}
\centering\includegraphics[width=80mm]{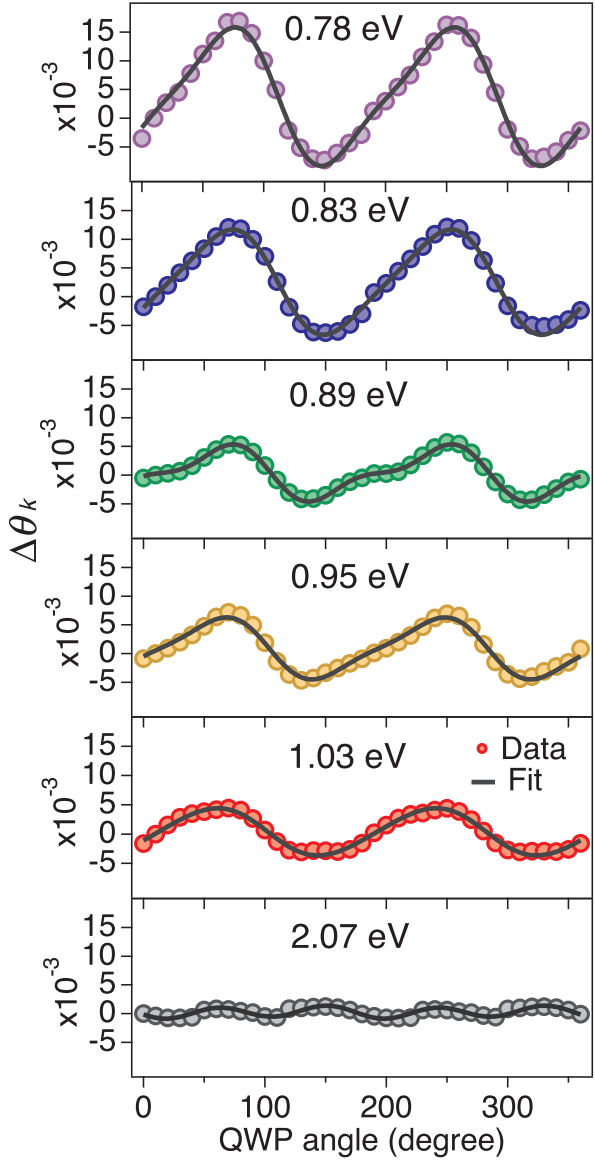}
\caption{QWP angle dependence of transient Kerr rotation. The peak intensity of the time resolved Kerr rotation is plotted for each photon energy. 
The closed circles represent the experimental data and the solid lines correspond to the fit using Eq. (\ref{eq:fit}). }
\label{fig:sinwave}
\end{figure}

Time resolved Kerr rotation signals ($\Delta\theta_{k}$) as a function of the angle of the QWP ($\alpha$) of 70$^{\circ}$ and 130$^{\circ}$ for the excitation wavelength of 1300 nm (0.954 eV) are presented in Fig.\ref{fig:setup}(b). The transient peak at time delay zero has the opposite sign and a different intensity level from the remainder of the spectra, depending on the pump helicity. Both $\alpha$ = 70$^{\circ}$ and 130$^{\circ}$ correspond to an elliptically polarized pump pulse and exhibit the largest positive and negative transient, respectively. The peak intensity of $\Delta\theta_{k}$ is plotted as a function of the QWP angle $\alpha$ in Fig.\ref{fig:sinwave} for different photon energies. The peak signals for the smaller photon energies shown in Fig.\ref{fig:sinwave} exhibit a significantly asymmetric sin(2$\alpha$) oscillation due to contributions from a smaller sin(4$\alpha$) oscillation. 
The stronger sin(2$\alpha$) oscillation observed in the present study for the CSL samples compared to the previous report \cite{mondal2018a} may be a consequence of the two-orders larger pump fluence used, which is expected to induce stronger SIFE arising from larger third-order susceptibility contributions. All the data were fit using Eq. (\ref{eq:fit}) to extract the contributions from the sin(2$\alpha$) and sin(4$\alpha$) components, corresponding to SIFE and SOKE signals, respectively \cite{wilks2004, popov1996}.

\begin{figure}
\centering\includegraphics[width=88mm]{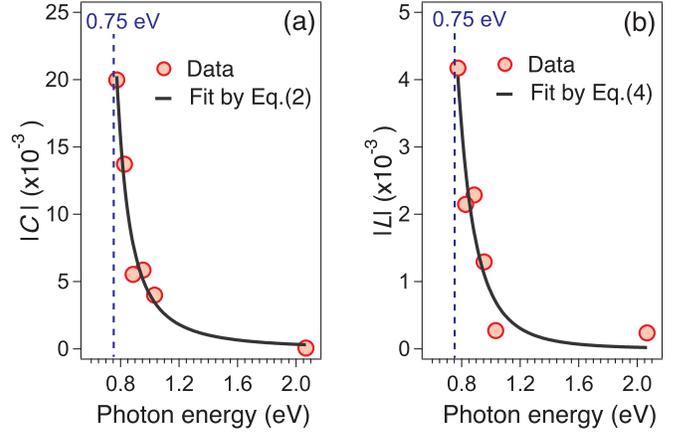}
\caption{Photon energy dependence of (a) $C$ and (b) $L$. The black solid curves are fits based on Eqs. (\ref{eq:ife}) and (\ref{eq:chi2}), respectively. 
The dashed line at 0.75 eV represents the upper bound photon energy required for exciting electrons from BVB bands to the Dirac bands. }
\label{fig:lc}
\end{figure}

\begin{equation}
\Delta \theta _{k}\ ( \alpha ) = C\sin( 2\alpha ) + L\sin( 4\alpha ) +D,
\label{eq:fit}
\end{equation}
where $D$ is a polarization independent background. 
The photon energy dependences of the $C$ (SIFE) and $L$ (SOKE) terms are shown in Fig.\ref{fig:lc}. As the photon energy decreased, the absolute values of $C$ and $L$ were found to sharply increase. 
The steeper increase in $L$ over $C$ may be a consequence of TI properties \cite{mondal2018a}.  
To investigate the sharp increases in the value of $L$ and $C$ shown in Fig. \ref{fig:lc}, the effect of the photon energy dependent susceptibility function on the SIFE and SOKE is examined, where second- and third-order nonlinear susceptibilities are expected to play major roles \cite{mondal2018}.
Based on the classical Drude-Lorentz model and the quantum mechanical description for the IFE \cite{battiato2014,berritta2016}, the response function for the induced magnetization (SIFE) for the off-resonant case can be simply expressed by \cite{robert2008}, 

\begin{equation}
 \chi ^{(3)}(\omega; \omega, \omega, -\omega) \propto \frac{1}{\omega^{2}( \omega -\omega_{0})},
\label{eq:ife}
\end{equation}
where $\omega _{0}$ is the resonant frequency of the system, which means that even though the SIFE is enhanced by being close to resonance, it does not vanish away from the resonance. 
Fig. \ref{fig:lc}(a) shows a fit of $|C|$ using Eq. (\ref{eq:ife}) with the fitting parameter of $\hbar \omega _{0}$ = 0.66 $\pm$ 0.03 eV.
Eq. (\ref{eq:ife}) was found to fit the experimental data ($|C|$) satisfactory. 
In contrast, for the SOKE case, it was assumed that the response function for $L$ was dominated by cascading $\chi ^{(2)} \cdot \chi ^{(2)}$ contributions \cite{mondal2018}. In quantum theory, the second-order nonlinear susceptibility for the off-resonant case can be expressed by permutation of indexes as \cite{robert2008,Bosshard95}:
\begin{equation}
\chi ^{(2)}(0; \omega, -\omega) = \chi ^{(2)}(-\omega; \omega, 0) \propto \frac{1}{( \omega -\omega_{0})^{2}},
\label{eq:or}
\end{equation}

where $\chi ^{(2)}(0; \omega, -\omega)$ represents optical rectification and $\chi ^{(2)}(-\omega; \omega, 0) $ represents the electro-optic effect. 
Thus, SOKE signal generated via cascaded optical rectification and the electro-optic effect, $\chi ^{(2)}(0; \omega, -\omega) \cdot \chi ^{( 2)}(-\omega; \omega, 0)$, is simply proportional to the square of the $\chi ^{(2)}$: 
\begin{equation}
\chi ^{( 2)} \cdotp \chi ^{( 2)} \propto \left[\frac{1}{( \omega - \omega _{0})^{2}}\right]^{2}.
\label{eq:chi2}
\end{equation}

The photon energy dependence of $|L|$ in Fig. \ref{fig:lc}(b) was well fit by Eq. (\ref{eq:chi2}) with $\hbar \omega _{0}$ = 0.62 $\pm$ 0.24 eV. 

\onecolumngrid
\begin{widetext}
\begin{figure}
\centering\includegraphics[width=160mm]{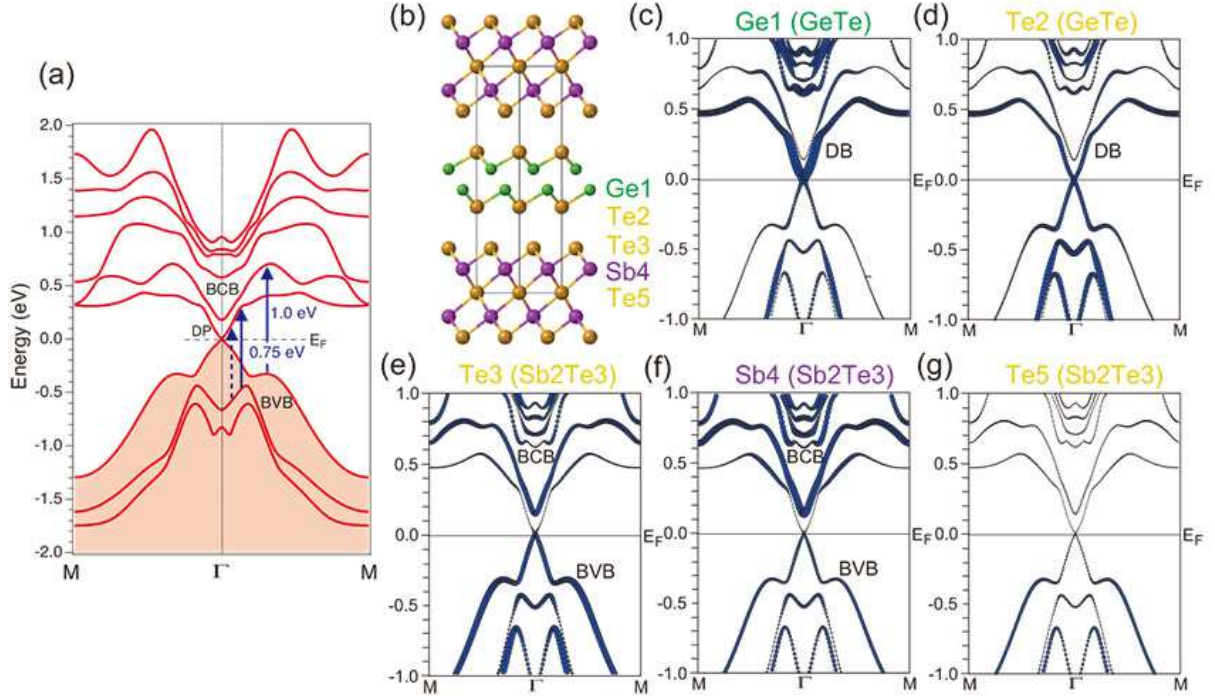}
\caption{(a) The electronic band structure of the chalcogenide superlattice, (GeTe)$_{2}$/Sb$_{2}$Te$_{3}$ inverted Petrov model, obtained by {\it ab initio} calculations \cite{saito2017}. 
The solid arrows indicate the transition energies necessary to excite electrons from BVB states to topological Dirac states above the Dirac point (DP) at 0.75 eV and that to BCB states at 1.0 eV. The dashed arrow represents a possible transition from BVB to near the DP ($\approx$0.65 eV). 
(b) The crystal structure of the inverted Petrov model. (c) - (g) Element projected band structures for each atomic layers shown in (b). The thicker bands represent stronger occupations. 
}
\label{fig:band}
\end{figure}
\end{widetext}
\twocolumngrid
The good fit results seen in Fig. 3 using Eqs. (\ref{eq:ife}) and (\ref{eq:chi2}) suggest that the increase in both $C$ (SIFE) and $L$ (SOKE) terms arises from possible resonant excitation at $\approx$0.60 -- 0.66 eV. 
According to the simuilated electronic band structure, one can see that photon energies less than 0.75 eV will promote electrons from the bulk valence band (BVB) to near the top of the Dirac band (DB), while for larger photon energies than 0.75 eV (e.g., 1.0 eV) a direct optical transition will occur from underlying BVB to the bulk conduction band (BCB) across the DB (see Fig. 4). 
It has been reported that in a magnetic topological insulator, the photo-induced current becomes larger when electrons are excited to topological surface states and reaches a maximum when electrons are excited from one surface band to another \cite{ogawa2016}. We propose that the steep increase in both $C$ and $L$ values toward lower photon energies corresponds to the tail of the resonant excitation from the BVB to around the Dirac point (DP) ($\approx$0.65 eV), as indicated by the dashed arrow in Fig. 4(a) \cite{Krizman2018}. Importantly, the steeper increase in $L$ relative to $C$ is expected to be a consequence of cascaded optical rectification and electro-optic effect, varying as $\sim \omega^{-4}$, where spatial inversion symmetry at the interfaces is broken by strain effects, leading a nonzero $\chi ^{(2)}$ \cite{mondal2018a}.
Note that the Dirac states in a CSL are either topological interface states \cite{Krizman2018} or topological bulk states \cite{tominaga2014}, as predicted by {\it ab initio} band structure simulations. 
Although further experiments using yet lower photon energies are required to fully understand the enhancement of the Kerr rotation in the topological chalcogenide superlattice, our results offer significant insight into the Dirac-cone oriented photon energy dependence of spin excitation in modern two dimensional materials. 

In summary, we have investigated the photon energy dependence of the Kerr rotation in a chalcogenide superlattice using infrared (1200$\sim$1600 nm) femtosecond laser pulses. The measured pump helicity dependence yields a contribution from both sin(2$\alpha$) and sin(4$\alpha$) components. Phenomenologically, the former contribution arises from the SIFE while the latter arises from SOKE. The photon energy dependence of the relative contributions, labeled $C$ and $L$ respectively, shows a monotonic increase with decreasing photon energy. 
The energy dependence of $C$ can be attributed to the response function $\chi ^{(3)}$, while that of $L$ can be attributed to the photon energy dependence of $\chi^{(2)}\cdot \chi^{(2)}$, where the response functions of $\chi ^{(2)}$ and $\chi ^{(3)}$ are derived from quantum theory. The sharp increase is attributed to resonant-like electronic excitation from BVB to Dirac states. 

\begin{acknowledgments}
This work was supported by CREST (Nos. JPMJCR14F1 and JPMJCR1875), JST, Japan and the Japan Society for the Promotion of Science (Grants-in-Aid for Scientific Research: 17H06088). 
We acknowledge Ms. R. Kondou for sample preparation.
\end{acknowledgments}


\section*{References}
\begin{thebibliography}{10}

\bibitem{simpson2011} R. E. Simpson, P. Fons, A. V. Kolobov, T. Fukaya, M. Krbal, T. Yagi, and J. Tominaga, Nat. Nanotech. \textbf{6}, 501 (2011).

\bibitem{lotnyk2017} A. Lotnyk, I. Hilmi, U. Ross, and B. Rauschenbach, Nano Res. \textbf{11}, 1676 (2017).

\bibitem{kim2017} J. Kim, J. Kim, Y. S. Song, R. Wu, S. H. Jhi, and N. Kioussis, Phys. Rev. B \textbf{96}, 235304 (2017).

\bibitem{makino2014} K. Makino, Y. Saito, P. Fons, A. V. Kolobov, T. Nakano, J. Tominaga, and M. Hase, Appl. Phys. Lett. \textbf{105}, 151902 (2014).

\bibitem{tominaga2014} J. Tominaga, A. V. Kolobov, P. Fons, T. Nakano, and S. Murakami, Adv. Mater. Interfaces, \textbf{1}, 1300027 (2014). 

\bibitem{Kellner2018}
J. Kellner, G. Bihlmayer, M. Liebmann, S. Otto, C. Pauly, J. E. Boschker, V. Bragaglia, S. Cecchi, R. N. Wang, V. L. Deringer, P. K\"{u}ppers, P. Bhaskar, E. Golias, J. S\'{a}nchez-Barriga, R. Dronskowski, T. Fauster, O. Rader, R. Calarco and M. Morgenstern, Commun. Phys., \textbf{5}, 1 (2018).

\bibitem{mclver2012} J. W. Mclver, D. Hsieh, H. Steinberg, P. Jarillo-Herrero, and N. Gedik, Nat. Nanotech. \textbf{7}, 96 (2012).

\bibitem{hsieh2011} D. Hsieh, F. Mahmood, J. W. Mclever, D. R. Gardner, Y. S. Lee, and N. Gedik, Phys. Rev. Lett. \textbf{107}, 077401 (2011).

\bibitem{boschinii2015} F. Boschini, M. Mansurova, G. Mussler, J. Kampmeier, D. Grutzmacher, L. Braun, F. Katmis, J. S. Moodera, C. Dallera, E. Carpene, C. Franz, M. Czerner, C. Heliger, T. Kampfrath, and M. Munzenberg, Sci. Rep. \textbf{5}, 15304 (2015).

\bibitem{Takeno2018} H. Takeno, S. Saito, and K. Mizuguchi, Sci. Rep. \textbf{8}, 15392 (2018). 


\bibitem{qiu2000} Z. Q. Qiu, and S. D. Bader, Rev. Sci. Instrum. \textbf{71}, 1243 (2000).

\bibitem{tse2010} W. K. Tse, and A. H. MacDonald, Phys. Rev. Lett. \textbf{105}, 057401 (2010).

\bibitem{wilks2004a} R. Wilks and R. J. Hicken, J. Appl. Phys. \textbf{95}, 7441 (2004).

\bibitem{wilks2003} R. Wilks, N. D. Hughes, and R. J. Hicken, J. Phys.: Condens. Matter \textbf{15}, 5129 (2003).

\bibitem{ziel1965} J. P. van der Ziel, P. S. Pershan, and L. D. Malmstrom, Phys. Rev. Lett. \textbf{15}, 190 (1965).

\bibitem{kimel2005} A. V. Kimel, A. Kirilyuk, P. A. Usachev, R. V. Pisarev, A. M. Balbashov, and Th. Rasing, Nature \textbf{435}, 655 (2005).

\bibitem{shen1984} Y. R. Shen, {\it The Principle of Nonlinear Optics} (Wiley, New York, 1984).

\bibitem{wilks2004} R. Wilks, and R. J. Hicken, J. Phys.: Condens. Matter \textbf{16}, 4607 (2004).

\bibitem{mondal2018} R. Mondal, Y. Saito, Y. Aihara, P. Fons, A. V. Kolobov, J. Tominaga, S. Murakami, and M. Hase, Sci. Rep. \textbf{8}, 3908 (2018).

\bibitem{mondal2018a} R. Mondal, Y. Aihara, Y. Saito, P. Fons, A. V. Kolobv, J. Tominaga, and M. Hase, Appl. Mater. Interfaces \textbf{10}, 26781 (2018).

\bibitem{mondal2018b} R. Mondal, A. Arai, Y. Saito, P. Fons, A. V. Kolobov, J. Tominaga, and M. Hase, Phys. Rev. B \textbf{97}, 144306 (2018).

\bibitem{Krizman2018} G. Krizman, B. A. Assaf, T. Phuphachong, G. Bauer, G. Springholz, G. Bastard, R. Ferreira, L. A. de Vaulchier, and Y. Guldner, Phys. Rev. B \textbf{98}, 075303 (2018).

\bibitem{Hase2012} M. Hase, M. Katsuragawa, A. M. Constantinescu, and H. Petek, Nat. Photon. \textbf{6}, 243 (2012).

\bibitem{Hase2015} M. Hase, P. Fons, K. Mitrofanov, A. V. Kolobov, and J. Tominaga, Nat. Commun. \textbf{6}, 8367 (2015).

\bibitem{lee2005} B. S. Lee and John R. Abelson, J. Appl. Phys. \textbf{97}, 093509 (2005).

\bibitem{saito2015} Y. Saito, P. Fons, A. V. Kolobov, and J. Tominaga, Phys. Status Solidi B \textbf{252}, 2151 (2015).

\bibitem{saito2016} Y. Saito, P. Fons, L. Bolotov, N. Miyata, A. V. Kolobov, and J. Tominaga, AIP Adv. \textbf{6}, 045220 (2016).

\bibitem{popov1996} S. Popov, Y. Svirko, and N. I. Zheludev, J. Opt. Soc. Am. B \textbf{13}, 2729 (1996).

\bibitem{battiato2014} M. Battiato, G. Barbalinardo, and P. M. Oppeneer, Phys. Rev. B \textbf{89}, 014413 (2014).

\bibitem{berritta2016} M. Berritta, R. Mondal, K. Carva, and P. M. Oppeneer, Phys. Rev. Lett. \textbf{117}, 137203 (2016).

\bibitem{robert2008} R. W. Boyd , {\it Nonlinear Optics} (Boston: Academic, 2008).

\bibitem{Bosshard95} Ch. Bosshard, R. Spreiter, M. Zgonik, and P. G\"{u}nter, Phys. Rev. Lett. \textbf{74}, 2816 (1995).


\bibitem{ogawa2016} N. Ogawa, R. Yoshimi, K. Yasuda, A. Tsukazaki, M. Kawasaki, and Y. Tokura, Nat. Commun. \textbf{7}, 12246 (2016).

\bibitem{saito2017} Y. Saito, K. Makino, P. Fons, A. V. Kolobov, and J. Tominaga, ACS Appl. Mater. Interfaces, \textbf{9}, 23918 (2017).


\end{thebibliography}
\end{document}